%% file: paper.tex
\documentclass[runningheads]{llncs}
\usepackage[T1]{fontenc}
\usepackage{graphicx}
\usepackage{times}
\usepackage{latexsym}
\usepackage{graphicx}
\usepackage{ulem}
\usepackage{amsmath}
\usepackage{amssymb}
\usepackage{graphicx}
\usepackage{float}
\usepackage{array}
\usepackage{multirow}
\usepackage[thinlines]{easytable}
\usepackage{changepage}
\usepackage{makecell}
\usepackage[symbol]{footmisc}

\begin{document}
\title{U-PET: MRI-based Dementia Detection with Joint Generation of Synthetic FDG-PET Images}
\titlerunning{U-PET: Dementia Detection and PET Generation}
\author{Marcel Kollovieh\inst{1}$^{,*}$
 \and Matthias Keicher\inst{1}$^{,*}$ \and Stephan Wunderlich\inst{3} \and Hendrik Burwinkel \inst{1} \and Thomas Wendler \inst{1} \and Nassir Navab \inst{1,2}}
\authorrunning{Kollovieh et al.}
\newcommand\asteriskfill{\leavevmode\xleaders\hbox{$\ast$}\hfill\kern0pt}
\institute{Computer Aided Medical Procedures, Technical University Munich, Germany
\and Computer Aided Medical Procedures, Johns Hopkins University, Baltimore, USA
\and Department of Radiology, Ludwig Maximilians University Munich, Munich, Germany}
\maketitle              % typeset the header of the contribution
\def\thefootnote{*}\footnotetext{Contributed equally.}
\begin{abstract}
Alzheimer’s disease (AD) is the most common cause of dementia. An early detection is crucial for slowing down the disease and mitigating risks related to the progression. While the combination of MRI and FDG-PET is the best image-based tool for diagnosis, FDG-PET is not always available. The reliable detection of Alzheimer’s disease with only MRI could be beneficial, especially in regions where FDG-PET might not be affordable for all patients. To this end, we propose a multi-task method based on U-Net that takes T1-weighted MR images as an input to generate synthetic FDG-PET images and classifies the dementia progression of the patient into cognitive normal (CN), cognitive impairment (MCI), and AD. The attention gates used in both task heads can visualize the most relevant parts of the brain, guiding the examiner and adding interpretability. Results show the successful generation of synthetic FDG-PET images and a performance increase in disease classification over the naive single-task baseline. 

\keywords{Dementia Detection  \and FDG-PET Generation \and Multi-tasking \and Domain transfer \and Attention Gate}
\end{abstract}

\input{chapters/introduction}
\input{chapters/related_work}
\input{chapters/methodology}

\input{chapters/experiments}

\input{chapters/results}
\input{chapters/conclusion}

\bibliographystyle{splncs04}
\bibliography{bibliography}
\end{document}

%% file: chapters/introduction.tex
\section{Introduction}
Alzheimer's disease (AD) is the most common cause of dementia. In contrast to many other pathologies, the death rate of AD is expected to rise \cite{Alzheimer_2020}. Yet, early detection enables slowing down the disease, reducing the AD-related death rate and mitigating related risks, by providing proper accompanying support and treatment. While there are various different tests for a doctor to diagnose the diseased brain, imaging can be a crucial factor. The most common and available technique to assess neuronal injury is Magnetic Resonance Imaging (MRI), however, 2-deoxy-2-[$^{18}$F]fluoro-D-glucose Positron Emission Tomography (FDG-PET, in this paper abbreviated solely as PET) is superior for this task \cite{lotan_brain_2020}. PET, nevertheless, is an expensive modality which is not as broadly available, in particular, in poorer world regions. Further, it requires the injection of radioactive glucose and thus an increased radiation burden to the patient.

There are several approaches to diagnose the AD given an MRI \cite{valliani2017deep,backstrom2018efficient,wen2020convolutional}. Even though they proved to be suitable for this task, there is no guarantee of their reliance, and they often lack explainability \cite{gradcam}. Following the WHO's recommendations, interpretability is paramount to create trust and help the medical doctor with the diagnosis, and avoid potentially severe consequences of wrong inference \cite{WHO_AI}.

Several methods have been proposed to tackle this issue by creating visual explanations in a classification task by highlighting the most crucial regions of the input \cite{cam,gradcam}. Another possibility would be to perform a modality transfer from MRI to PET images, which are more suitable for the medical doctor to diagnose the patient as proposed in \cite{sikka2018mri,jung2018inferring,sun2019dual}. Moreover, if reliable, inferring PET images from MRI would make PET-based diagnostics more widely available in poor world regions. At the same time, the radioactive tracer needed to acquire PETs could be left off thus reducing radiation burden and costs.

In this work, we investigate the feasibility of visualizing techniques and modality transfers using MRI for the purpose of a better interpretable AI-powered AD classification. Our contributions can be summarized as follows:

\begin{itemize}
    \item We propose \textit{U-PET}, a neural network suitable for dementia detection and joint synthetic PET generation.
    \item We show its superior performance compared to other models by quantitatively evaluating its diagnosis predictions using several metrics.
    \item We evaluate the effect of the auxiliary task, namely, the synthetic PET generation.
    \item We provide a degree of interpretability to our method by using attention maps for both tasks which provide important information of its decision process.
\end{itemize}

%% file: chapters/related_work.tex
\subsubsection{Related Work}
There are several works based on machine learning related to AD research, most starting from MRI. While many approaches conduct a binary classification to distinguish between AD and cognitive normal (CN) images. Some also include other groups such as mild cognitive impairment (MCI) and either perform a different binary or a multiclass classification \cite{wen2020convolutional}. Even though MRIs are 3D volumes, there are different ways they are handled. While some directly use the 3D volumes \cite{backstrom2018efficient}, other slice the volume and either use one slice \cite{valliani2017deep}, or a majority voting system \cite{wen2020convolutional} to infer the diagnose. Such methods reach balanced accuracy ranging 0.64-0.89 depending on the dataset in the task of static AD/CN classification, yet none analyse in depth their explainability.

When using PET images for the classifications, easier methods such as support vector machines (SVM) seem to be sufficient if they are applied on segmented brains \cite{gray2011regional}, or in combination with principal component analysis (PCA) \cite{lopez2011principal}. However, neural networks are also used \cite{liu2018classification} getting an area under the receiver-operator curve (AUC) of up to 0.95. Yet, again, they do not offer means for explaining the network's reasoning.

Modality transfer was also tested using Pix2Pix \cite{jung2018inferring} and more advanced conditional models \cite{sun2019dual}. The goal there is to generate PET images from MRI to aid physicians with "an-easier-to-interpret" visualization and avoid the downsides of PET. Furthermore, classification conducted on synthetic PET images obtained from MRI yielded a statistically significant improved accuracy of 0.74 vs. 0.70 when compared to classification of MRI only \cite{sikka2018mri}. These results encourage further analysis as possibly MRI could be further exploited by reshaping its information as PET. While \cite{sikka2018mri} used a sequential approach, we propose a multitask approach.

In this work, we combine neural-network classification of MRI with strategies to generate PET. We hypothesize that by generating a synthetic PET from MRI using the same neural network used for classification, we can both boost performance, while providing easier to interpret images for physicians. Additionally, by adding attention-mechanisms we claim we are making a step towards explainability in MRI-based AD prediction.

%% file: chapters/methodology.tex
\section{Methodology}
In this section, we describe our model which is leveraged for classification and PET generation tasks. Further, we give details on its interpretability.
\subsection{Multitask learning via U-Net}
\label{section:unet}
\begin{figure*}[h]
  \centering
  \includegraphics[width=0.9\textwidth]{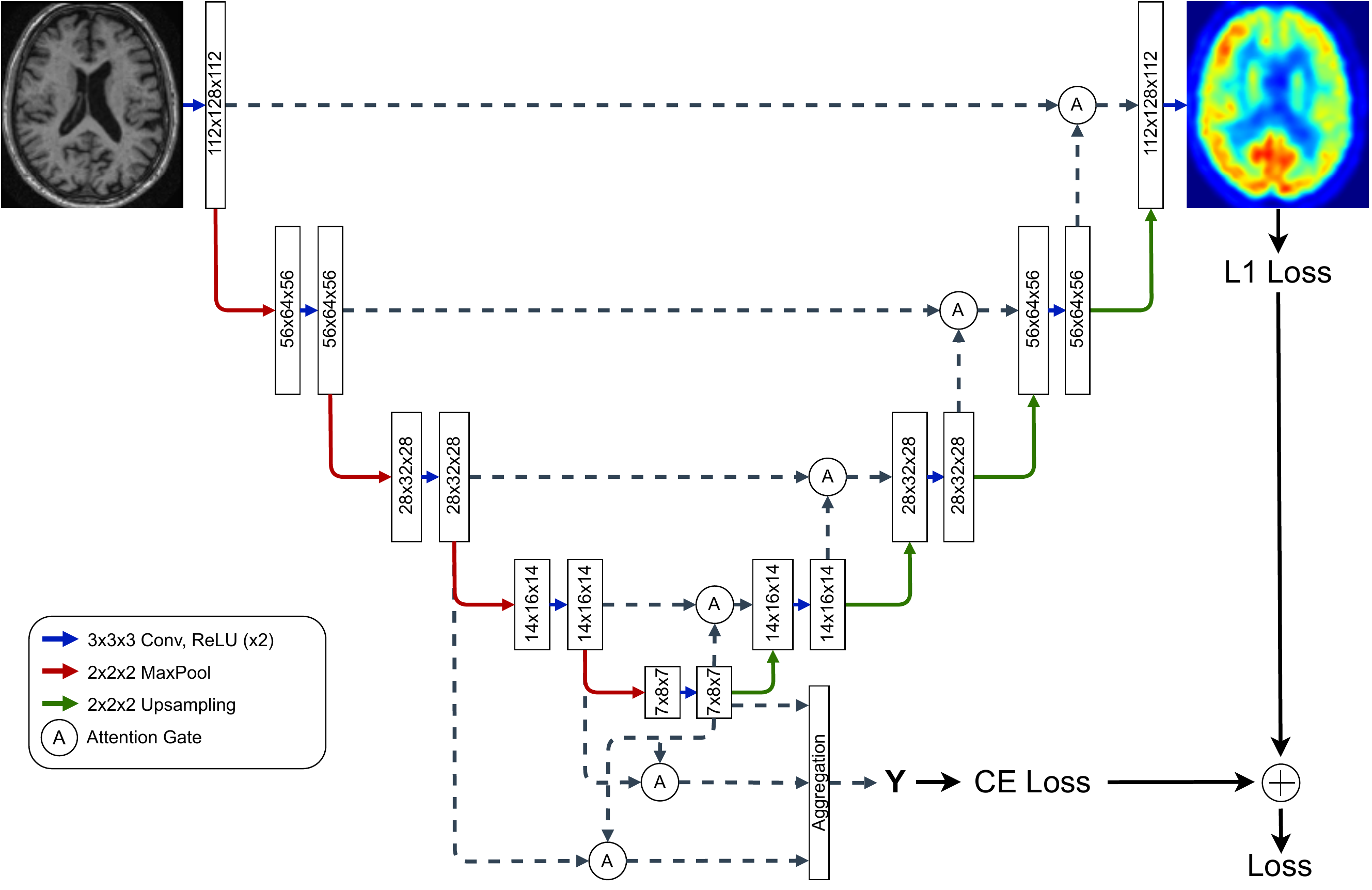}
  \caption{Our proposed \textit{U-PET} focuses on the classification with joint PET generation. The model is based on \cite{oktay2018attention}. We add linear layers to the bottleneck to infer the diagnosis $Y$, thus allowing the network to share features of the classification and generation task. The skip connections are forwarded through attention gates. Instead of only using the final bottleneck output, we aggregate the two previous level's final feature maps and forward them through attention gates to infer the label similarly to \cite{schlemper2018attention}. }
  \label{fig:attunet}
\end{figure*}

The main goal of this work is to provide interpretability to the classification of MRI. While Grad-CAM \cite{gradcam} visualizes the decision process of the neural network, our approach aims at generating a synthetic PET together with the diagnosis. For this, first, we devise a U-Net-based architecture - which we named \textit{U-PET} - to both convert MRI to synthetic PET (one head) and classify the input MRI between AD, MCI and CN.

\textit{U-Net} is a widely used network for segmentation which consists of an encoder and decoder, and has lateral skip connections between the corresponding levels of the encoder and decoder \cite{ronneberger2015u}. There have been several modifications such as the self-adapting \textit{nnU-Net} \cite{isensee2019automated,isensee2018nnu}. Oktay et al. propose the \textit{Attention U-Net} which applies attention gates on the skip connections to improve the foreground pixel sensitivity. Furthermore, the attention maps can be visualized to highlight important regions in the input \cite{oktay2018attention}.

Our \textit{U-PET} is based on the \textit{Attention U-Net} \cite{oktay2018attention} and extended to infer PET images from MRIs. Further, we introduce a more advanced scheme for the classification. Instead of using only the bottleneck layer, we aggregate it with the feature maps from the two previous levels to infer the prediction as proposed by \cite{schlemper2018attention}. This aggregation scheme allows the model to use features from the coarse and fine scales.

The output of an attention gate is derived by an element-wise multiplication of the input $x$ and the attention coefficients $\alpha$. Formally written, the output at position $i$ of channel $c$ corresponds to $\hat{x}_{i,c}=x_{i,c}\cdot\alpha_i$.
To infer the attention coefficients additive attention is used: 
\begin{align*}
   \alpha_i=\sigma_2(\psi^T(\sigma_1(W_x^Tx_i+W_g^Tg_i+b_g)+b_\psi))
\end{align*}
where $\psi$, $W_x$ and $W_g$ denote linear transformations, $b_g$ and $b_\psi$ biases, $g_i$ a gating signal, $\sigma_1(\cdot)$ a rectified linear unit (ReLU) and $\sigma_2(\cdot)$ a sigmoid function.  \cite{oktay2018attention} \\

We use attention gates for the skip connections, the input is the last set of activation maps from the corresponding encoder level the gating signal is the last activation map from the previous (coarser) level in the decoder. Furthermore, we also use attention gates for the the classification where the bottleneck feature maps are taken as gating signals and the feature maps from the previous levels as input. The different outputs are aggregated by using the mean as in \cite{schlemper2018attention}. The architecture is shown in Fig.~\ref{fig:attunet}. Using this approach, we do not only compute classification and a synthetic PET image but, in addition, we can visualize the attention maps, both, for the classification and for the PET generation.

%% file: chapters/experiments.tex
\section{Experimental Setup}
\input{chapters/dataset}

\subsection{PET Generation and AD Detection}
\subsubsection{Experiments}
To compare our model we use several different baselines. For the classification, we use a \textit{DenseNet121} \cite{huang2017densely} from the MONAI library \cite{the_monai_consortium_2020_4323059} and for the modality transfer (i.e. synthetic PET generation), we use a \textit{pix2pix} with a \texttt{unet128} as generator \cite{isola2017image}. Furthermore, we test the importance of the attention gates by comparing it with a modification which has no attention mechanism \textit{U-PET (no att.)} and is based on the MONAI library\cite{the_monai_consortium_2020_4323059}. Finally, to test the importance of the joint PET generation for the classification, we train our \textit{U-PET} without it (\textit{U-PET (no PET)}). The classification performance is compared using the accuracy, F1 score (Macro) and AUC for each class. The modality transfer is evaluated using the mean absolute error (MAE).
\subsubsection{Implementation details}
The cost function for the \textit{DenseNet121} and \textit{U-PET-N} is the cross-entropy loss. The multi-task methods, namely the \textit{U-PET (no att.)} and \textit{U-PET} are trained on the sum of the cross-entropy and L1 loss with deep-supervision \cite{lee2015deeply}. The L1 loss is reduced by averaging over the voxels. In total we perform three runs for each model over 80 epochs using the Adam optimizer with a learning rate of 0.001 and a batch size of 4. We perform early stopping using the F1 score (Macro) on the validation set. For the implementation we use MONAI (0.4.0), PyTorch (1.7.0) and Pytorch Lightning (1.1.1).

%% file: chapters/dataset.tex
\subsection{ADNI Dataset}
The Alzheimer’s Disease Neuroimaging Initiative (ADNI) provides MRI and PET volumes of subjects \cite{ADNI}. These volumes are diagnosed as CN, AD or MCI.

To bring the data into an organized structure, it is converted into the Brain Imaging Data Structure (BIDS) format \cite{gorgolewski2016brain} using Clinica \cite{clinica,samper2018reproducible}. While neural networks perform well with minimal preprocessing, here the performance is drastically improved by several normalization steps on the MRIs \cite{backstrom2018efficient}. The influence of a minimalistic and an extensive preprocessing was studied by Wen et al. \cite{wen2020convolutional}.

To train and validate our approach, we need paired T1-weighted MRIs and PET images. Therefore, we collect the MRIs preprocessed by GradWarp, B1 correction and N3 bias field correction, and PET images which were co-registered and averaged \cite{ADNI}. Corresponding pairs are determined by matching subject- and session-ID. Both, the MRIs and the PETs are converted into the BIDS format.

\subsubsection{MRI preprocessing}
Then, we apply the \texttt{t1-volume} pipeline of Clinica \cite{clinica,samper2018reproducible} which runs several procedures from the SPM software to preprocess the MRIs. First, the \texttt{Segmentation} procedure performs \textit{unified segmentation}, which applies tissue segmentation, bias correction and spatial normalization \cite{ashburner2005unified}. Then, \texttt{Run Dartel} runs the DARTEL algorithm for diffeomorphic image registration, creating a group template and computing deformation fields from the previously obtained tissue maps in native space \cite{ashburner2007fast}. This template is needed to transform the MRIs from native to MNI space using the \texttt{Dartel2MNI} procedure \cite{ashburner2007fast}. After the pipeline is finished, we perform z-score normalization. Finally, the MRIs are cropped to a size of $112\times128\times112$ with voxel sizes of $1.5\times1.5\times1.5\text{mm}^3$.

\subsubsection{PET preprocessing}
To preprocess and register the PET images, we use the Clinica pipeline \texttt{pet-volume} \cite{clinica,samper2018reproducible}. First, the PET images are registered into the corresponding MRI's native space using the SPM software. Then, the PET images are spatially normalized into the MNI space using the DARTEL deformation model of SPM \cite{ashburner2007fast}, which was obtained in the $\texttt{t1-volume}$ pipeline. Furthermore, the intensity is normalized using the average PET uptake in a reference region (pons).\\

Using this procedure, we obtain 2616 MRIs from 636 subjects, 787 are labelled as CN, 757 as AD and 1072 as MCI. Out of these MRIs, 1000 have a matching PET image. We split the data as shown into a train (444 subjects, 1805 MRIs, and 686 PETs), validation (96, 408 and 154, respectively) and test (96, 403 and 160, respectively).
\iffalse
\begin{table}[h!]
    \caption{Split of the MRIs and PET images into the three different sets. The split is performed on a subject level to avoid data leakage.}

    \centering
    \begin{tabular}{c|c|c|c}
         & Train & Validation & Test\\
         \hline
         Subjects & 444 & 96 & 96 \\
         \hline
         MRIs  & 1805 & 408 & 403\\
         \hline
         PET images  &  686 & 154 & 160
    \end{tabular}
    \label{tab:split2}
\end{table}
\fi

%% file: chapters/results.tex
\section{Results and Discussion}
\subsubsection{Quantitative Results}
The classification performance is shown in table \ref{tab:dynunet}. The \textit{U-PET} performed best in our tests regarding accuracy and F1 score (Macro), while the \textit{DenseNet121} was the worst. The \textit{U-PET (no att.)} has a better performance than the \textit{DenseNet121}, but still worse than the \textit{U-PET}. Furthermore, the \textit{U-PET} without the reconstruction performs significantly worse, having a similar performance as the \textit{DenseNet121}. The superior performance of the \textit{U-PET} and \textit{U-PET (no att.)} implies the advantage of the reconstruction of the PET images for the classification, forcing the model to share its bottleneck weights. In the AUC metric, it can be seen that all the models struggle with the class MCI. Our best model, the \textit{U-PET} reaches an accuracy of 57.2\% while \cite{valliani2017deep} reach 56.8\%. It should be noted that while we split the data into a train, validation and test set, \cite{valliani2017deep} split into train and test, unknown to us, limiting the comparison. On the validation set, all our models achieve accuracies above 60\%. Furthermore, their best accuracy with a non-pretrained model is 50.9\% which matches our \textit{U-PET (no PET)}.

\begin{figure}
  \centering
  \includegraphics[width=1. \textwidth]{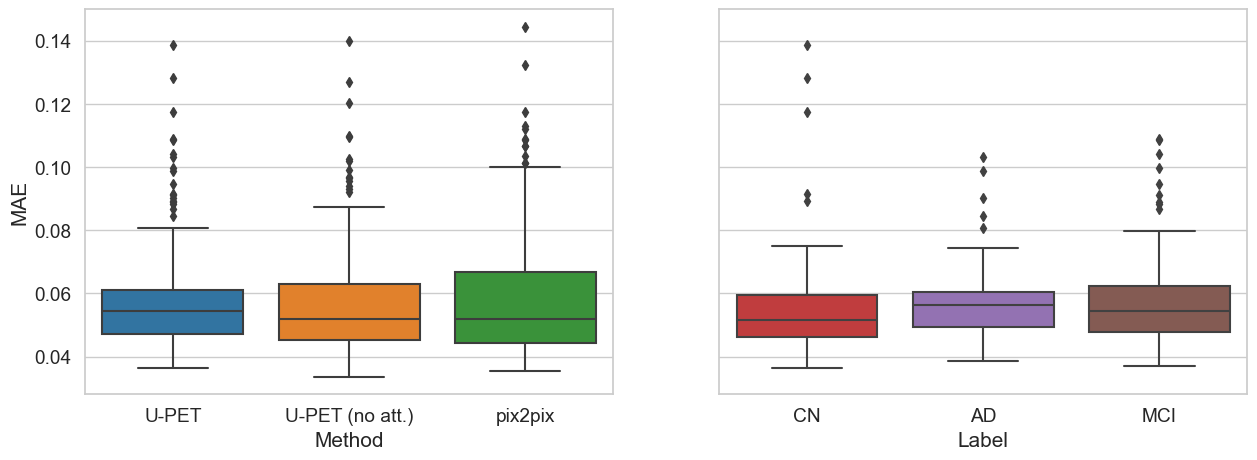}
  \caption{Left: MAE distribution of PET predictions on the test set for the different approaches. The corresponding means are \textit{U-PET}: 0.0586, \textit{U-PET (no att.)}: 0.0575 and \textit{pix2pix}: 0.0592. Right: MAE distribution of PET predictions on the test set for the \textit{U-PET} separated by labels. }
  \label{fig:violine}
\end{figure}

\begin{table*}
    \caption{Comparison of the \textit{DenseNet121}, \textit{U-PET (no att.)}, our \textit{U-PET (no PET)} and \textit{U-PET}. The metrics are computed on the test set.}

    \centering
    \begin{tabular}{l|c|c|c}
       & Accuracy & F1 (Macro) &  \makecell{AUC \,\\CN \qquad\quad\qquad AD \quad\qquad\qquad MCI} \\
       \hline
    \textit{DenseNet121} & 0.504 $\pm$ 0.024 & 0.506 $\pm$ 0.017 & 0.671 $\pm$ 0.017 \hspace{1mm} 0.653 $\pm$ 0.019 \hspace{1mm} 0.533 $\pm$ 0.048\\ \hline
    \textit{U-PET (no att.)} & 0.523 $\pm$ 0.020 & 0.529 $\pm$ 0.024 & 0.683 $\pm$ 0.054 \hspace{1mm} 0.693 $\pm$ 0.010 \hspace{1mm} 0.540 $\pm$ 0.010\\
    \hline
    \textit{U-PET (no PET)} & 0.509 $\pm$ 0.047 & 0.507 $\pm$ 0.052 & 0.711 $\pm$ 0.046 \hspace{1mm} \textbf{0.703 $\pm$ 0.015} \hspace{1mm} 0.527 $\pm$ 0.033\\
    \hline
    \textit{U-PET} &\textbf{0.572 $\pm$ 0.020} & \textbf{0.569 $\pm$ 0.017} & \textbf{0.752} $\pm$ \textbf{0.033} \hspace{1mm} 0.680 $\pm$ 0.017 \hspace{1mm} \textbf{0.597} $\pm$ \textbf{0.023}\\ 
    \end{tabular}
    \label{tab:dynunet}
\end{table*}
 Looking at the modality transfer, we can see in Fig.~\ref{fig:violine} that there is not a significant difference in the models' performance. While here the best model is the \textit{U-PET (no att.)} with an average MAE of 0.0575 the worst is the \textit{pix2pix} with an average MAE of 0.0592. The \textit{U-PET} performs between these two with an average MAE of 0.0586. Furthermore, it can be seen that the variance of the \textit{U-PET} is the lowest, followed by the \textit{U-PET (no att.)}, while the \textit{pix2pix} has the highest variance. Furthermore, there is no substantial performance gap for the different classes, showing that the \textit{U-PET} is able to consistently create PET predictions for all types of diagnosis.

\begin{figure*}[h!]
\centering
\begin{tabular}{ccccc}
  \includegraphics[width=0.177\textwidth]{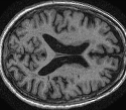} &    \includegraphics[width=0.177\textwidth]{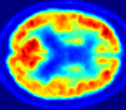} & \includegraphics[width=0.177\textwidth]{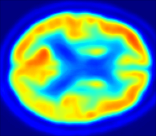} &
  \includegraphics[width=0.177\textwidth]{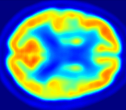} &    \includegraphics[width=0.177\textwidth]{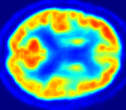} \\
  \includegraphics[width=0.177\textwidth]{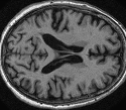} &    \includegraphics[width=0.177\textwidth]{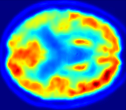} & \includegraphics[width=0.177\textwidth]{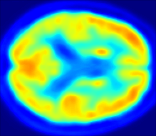} &
  \includegraphics[width=0.177\textwidth]{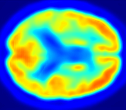} &    \includegraphics[width=0.177\textwidth]{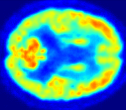} \\
  \includegraphics[width=0.177\textwidth]{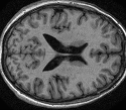} &    \includegraphics[width=0.177\textwidth]{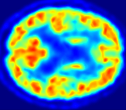} & \includegraphics[width=0.177\textwidth]{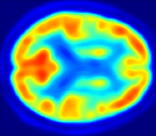} &
  \includegraphics[width=0.177\textwidth]{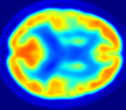} &    \includegraphics[width=0.177\textwidth]{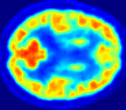} \\
  \includegraphics[width=0.177\textwidth]{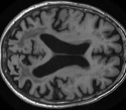} &    \includegraphics[width=0.177\textwidth]{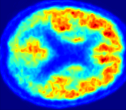} & \includegraphics[width=0.177\textwidth]{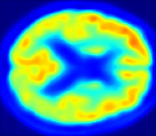} &
  \includegraphics[width=0.177\textwidth]{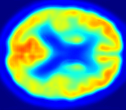} &    \includegraphics[width=0.177\textwidth]{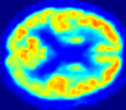} \\
  MRI & PET & \textit{U-PET} & \textit{U-PET (no att.)} & \textit{pix2pix}
\end{tabular}
\caption{Examples of synthetic PET images. In the first column the input MRI is shown, in the second column the ground truth PET. The last three columns are the predictions of the three different models. While the first three rows are good examples, the last row is a bad prediction.}
\label{fig:pets}
\end{figure*}
\subsubsection{Qualitative Results}
Looking at sample synthetic PET images in Fig.~\ref{fig:pets}, it can be seen that all of the models are able to make reasonable predictions. The outputs of the \textit{U-PET} and \textit{U-PET (no att.)} look very similar and are smoothed compared to the input. The \textit{pix2pix} predicts more details, which seem to be important for the discriminator. However, even though this model has finer predictions, the intensity values are worse compared to the \textit{U-PET} and \textit{U-PET (no att.)} as seen in Fig.~\ref{fig:violine}; an example is given in Fig.~\ref{fig:pets} in the second row.

From a clinical perspective, the synthetic generated PET show a smoother version of the real PET images but keep the same pattern of FDG uptake, especially, hypometabolism in the respective brain areas. This backs up our hypothesis, that MRI contains information that correlates with the functional information of PET images. Moreover, the synthetic PET could be used as complementary visualization for physicians beyond the task of classification.

\begin{figure*}[h!]
\centering
\begin{tabular}{lccccc}
  \rotatebox[origin=l]{90}{\hspace{4mm} CN}&\includegraphics[width=0.15\textwidth]{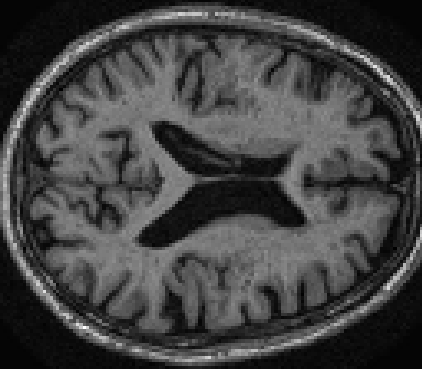} &    \includegraphics[width=0.15\textwidth]{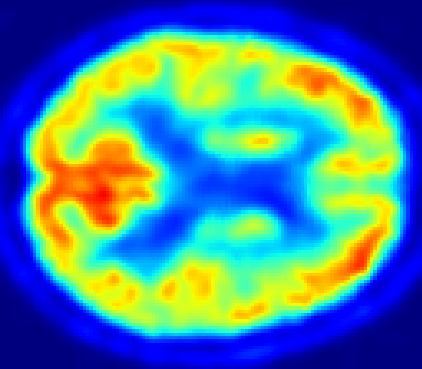} & \includegraphics[width=0.15\textwidth]{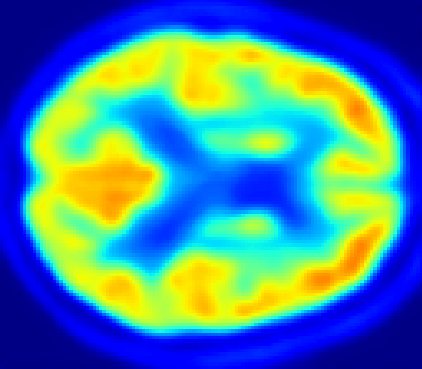} &
  \includegraphics[width=0.15\textwidth]{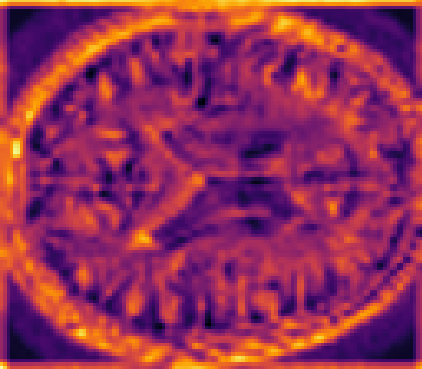} & 
  \includegraphics[width=0.15\textwidth]{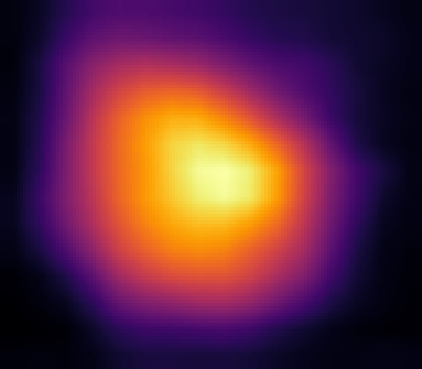}     \\
  \rotatebox[origin=l]{90}{ \hspace{3mm}MCI}&\includegraphics[width=0.15\textwidth]{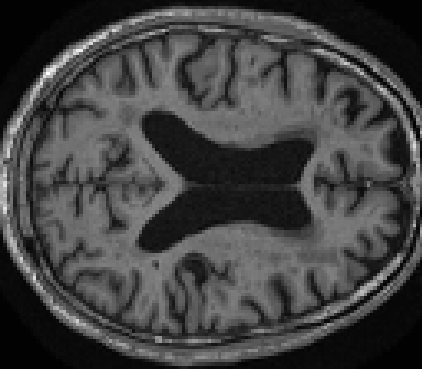} &    \includegraphics[width=0.15\textwidth]{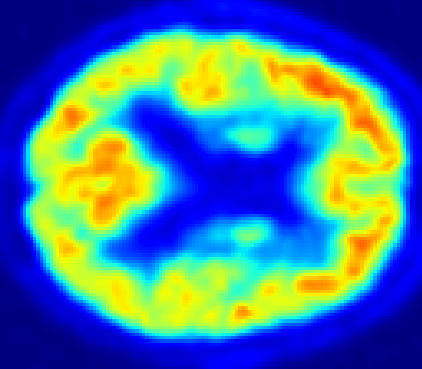} & \includegraphics[width=0.15\textwidth]{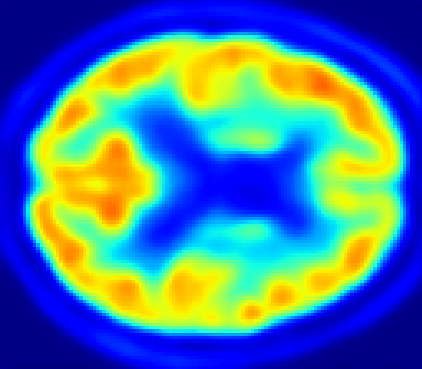} &
  \includegraphics[width=0.15\textwidth]{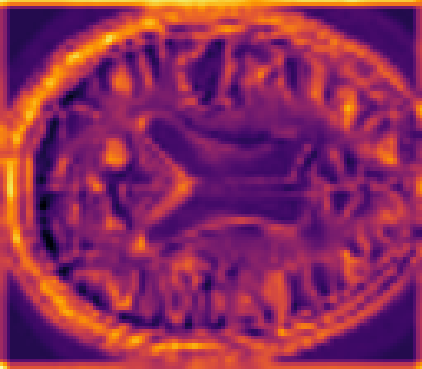} & 
  \includegraphics[width=0.15\textwidth]{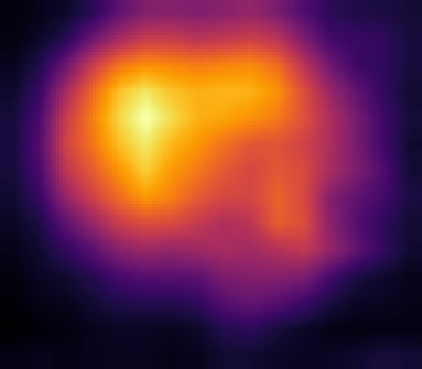}  \\
    \rotatebox[origin=l]{90}{\hspace{4mm} AD}&\includegraphics[width=0.15\textwidth]{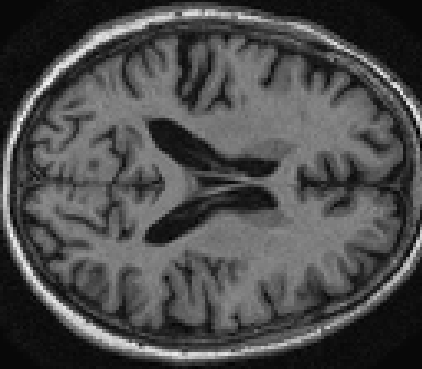} &    \includegraphics[width=0.15\textwidth]{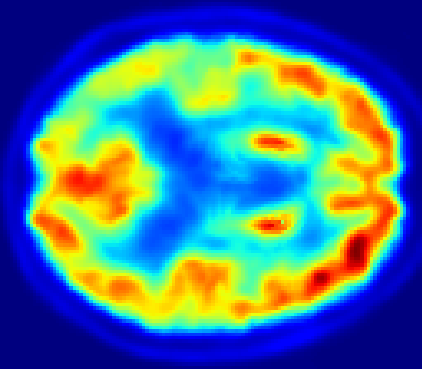} & \includegraphics[width=0.15\textwidth]{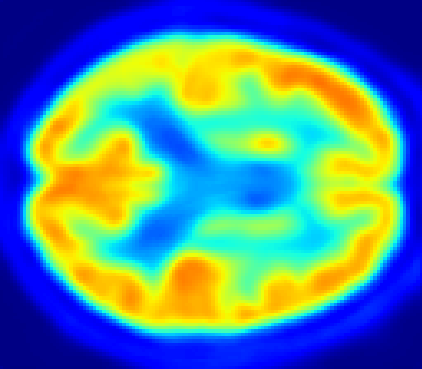} &
  \includegraphics[width=0.15\textwidth]{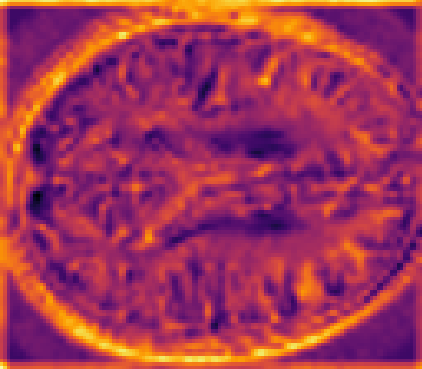} & 
  \includegraphics[width=0.15\textwidth]{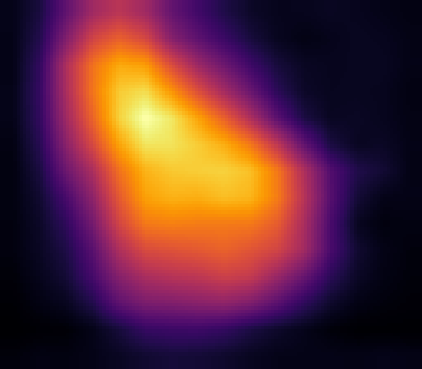}  \\
  & MRI & PET & Synthetic PET & AttMap skip & AttMap
\end{tabular}
\caption{Examples of synthetic PET images of the \textit{U-PET} and the corresponding attentions maps (AttMap). In the first column, the input MRI is shown, in the second and third columns, the ground truth and synthetic PET. The fourth column depicts the attention maps of the skip connection at the highest level. Column five shows the second attention map from the classification mechanism. The first row is a CN, the second row a MCI and the last row an AD subject.}
\label{fig:attmaps}
\end{figure*} 

In Fig.~\ref{fig:attmaps}, the attention maps of the \textit{U-PET} are shown for multiple examples. On the one hand, the attention map of the skip connection (AttMap skip) highlights details of the brain structure, which is in agreement with the assumption that the network has to transform the specific structures of the MRI scan into the PET modality. On the other hand, the attention gates used for the classification task rather focus on more specific regions. One can observe a trend that the attention maps used for classification tend to highlight regions which have a low uptake in the PET. Since areas with a lower uptake in the PET correspond to areas with lower functional activity (hypometabolism), this focus area of the network seems reasonable with respect to the classification task.

%% file: chapters/conclusion.tex
\section{Conclusion}
In this work, we investigate our multitask approach called \textit{U-PET}, combining classification and modality transfer from MRI to PET for the early detection of AD. The classification of \textit{U-PET} outperforms the compared baseline models while having a comparable PET generation performance. We also show that the classification performance can benefit from a joint generation of PET images. Our intuition for this is that the simultaneous generation of functional PET images guides the network to focus on more informative structures in the anatomical brain MRI with respect to the classification task. Moreover, our synthetic PET images overall show reasonable areas of hypometabolism compared to the real counterparts, indicating clinical meaningfulness.

As a future work, it would be highly interesting to evaluate prospectively how far the synthetic PET images differ from real ones, and if an expert-based analysis of the synthetic PET images yields a consistent classification result of \textit{U-PET}.